\documentclass[sigconf]{acmart}
\acmSubmissionID{128}

\usepackage[inline]{enumitem} 
\usepackage{xcolor} 
\usepackage{calc} 
\usepackage[skip=2pt]{caption} 

\usepackage[para]{footmisc}  
\usepackage{tabularx} 

\usepackage{tikz} 

\usepackage{acmart-taps}

\AtBeginDocument{%
  }

\newcommand{\OurParagraph}[1]{\vspace*{0.8mm}\noindent\textbf{#1}}

\definecolor{shadecolor}{rgb}{0.95,0.95,1}

\aptLtoXcmd{\newenvironment{bluebox}{\begin{xmlelement*}{aptdispbox}\XMLaddatt{style}{width: 900px;  background-color: \#f2f2ff; border: 1px solid \#000000; padding: 10px; }}{\end{xmlelement*}}}{\newsavebox{\blueboxcontent}
\newenvironment{bluebox}
  {\begin{lrbox}{\blueboxcontent}
   \begin{minipage}{\linewidth-15pt}
  }
  {
   \end{minipage}
   \end{lrbox}
   \begin{center}
   \begin{tikzpicture}
   \node[rounded corners=4pt, draw=black, line width=1.3pt,
         fill=shadecolor, inner sep=6pt]
   {\usebox{\blueboxcontent}};
   \end{tikzpicture}
   \end{center}
  }}

\begin{document}


\author{Jasmin Kareem}
\orcid{0009-0007-6398-1773}
\affiliation{%
  \institution{University of Amsterdam}
  \city{Amsterdam}
  \country{The Netherlands}
}
\affiliation{%
  \institution{Jheronimus Academy of Data Science}
  \city{'s-Hertogenbosch}
  \country{The Netherlands}
}
\email{j.kareem@tue.nl}

\author{Siddharth Mehrotra}
\orcid{0000-0002-2067-3451}
\affiliation{%
  \institution{Birla Institute of Technology \& Science, Pilani}
  \city{Pilani}
  \country{India}
}
\email{siddharth.mehrotra@pilani.bits-pilani.ac.in}

\author{Martijn C. Willemsen}
\orcid{0000-0001-5908-9511}
\affiliation{%
  \institution{Jheronimus Academy of Data Science}
  \city{'s-Hertogenbosch}
  \country{The Netherlands}
}
\affiliation{%
  \institution{Eindhoven University of Technology}
  \city{Eindhoven}
  \country{The Netherlands}
}
\email{m.c.willemsen@tue.nl}

\author{Maarten de Rijke}
\orcid{}
\affiliation{%
  \institution{University of Amsterdam}
  \city{Amsterdam}
  \country{The Netherlands}
}
\email{m.derijke@uva.nl}

\renewcommand{\shortauthors}{Kareem et al.}



\setcopyright{rightsretained}
\copyrightyear{2026}
\acmYear{2026}
\setcopyright{cc}
\setcctype{by}
\acmConference[RecSys '26]{20th ACM Conference on Recommender Systems}{September 27-October 02, 2026}{Minneapolis, MN, USA}
\acmBooktitle{20th ACM Conference on Recommender Systems (RecSys '26), September 27-October 02, 2026, Minneapolis, MN, USA}
\acmDOI{10.1145/3773078.3831744}
\acmISBN{979-8-4007-2284-4/2026/09}




\title[Do We Care About Personalization and Explainability?]{Do We Care About Personalization and Explainability?\\ An Interview Study with News Recommendation Engineers}

\begin{abstract}
  Research on explainability in recommender systems largely centers on end users, overlooking the perspectives of those who build and maintain these systems and their potential use cases such as model debugging. In this study, we examine how news engineers and related technical stakeholders perceive and implement personalization and explainability in practice. We conducted 15 semi-structured interviews across nine news organizations, spanning diverse regions in both public and private sectors, to investigate the challenges and motivations shaping their approaches. Our findings reveal that personalization is not always a straightforward or desirable choice for news organizations, as concerns around user tracking, editorial control, and resource constraints often limit its adoption. Even among organizations implementing personalized news recommender systems in production, explainability is rarely prioritized, with day-to-day operational demands frequently taking precedence over longer-term transparency goals.  Definitions of explainability vary widely across organizations, though some demonstrate promising internal practices and visualization tools that facilitate communication between engineering teams and newsrooms. Based on our analysis, we provide actionable and practical guidelines for news engineers and researchers on how to adopt explainability methods within a news personalization pipeline. 
\end{abstract}

\maketitle


\section{Introduction}
Personalization in news recommender systems has received growing attention from both academia and industry due to its potential to connect diverse audiences with relevant news content at the right time \cite{Wu2023Personalized}. Reflecting this trend, research on algorithmic approaches to news recommender systems has expanded rapidly. This body of work has largely focused on optimizing technical performance, while more user-centered and ``value-aware'' perspectives remain comparatively underexplored \cite{Bauer2024Where}. This imbalance is concerning, as news recommender systems play a central role in shaping public discourse and supporting democratic processes \cite{helberger2021democratic}. 

\OurParagraph{Values in news recommender systems (NRSs).}
A growing line of research emphasizes the importance of embedding societal and editorial values into news recommendation \cite{Kruse2024RecSys,panteli2019recommendation}. E.g., \citet{Vrijenhoek2022RADio} demonstrate how normative values can be operationalized in news recommender systems. Building on this, \citet{Lu2020beyond} incorporate editorial values into personalized news recommendations and evaluate their approach through user studies. 

\OurParagraph{News recommender systems and explainability.}
Next to diversity, transparency has emerged as a key value in the context of news recommendation. Understanding the mechanisms behind recommendation algorithms is essential for maintaining trust in journalism and news organizations \cite{diakopoulos2017algorithmic, moller2024designing}. Explainability techniques are commonly proposed as a means to increase transparency, traditionally focusing on helping readers understand why a particular news article was recommended to them \cite{ter2017news,explainingMoller2025,Liu2024Topic}. However, explainability encompasses multiple definitions and goals; \citet{tintarev2015explaining} identify several explanatory goals, including transparency, which aims to explain how a system works, and scrutability, which allows users to indicate when the system is wrong. For example, \citet{Sullivan2019reading} show that explanations grounded in user profiles can support readers’ sense of self-understanding and agency. However, end-users are not the only target group that require transparency, as there are multiple stakeholders that could benefit from explanations~\cite{lucic2021multistakeholder}. 

\OurParagraph{Explainability for news engineers.}
Recent work has begun to explore alternative roles for explainability, particularly as an internal tool for news organizations. \citet{cools2025navigating} examine the internal and external uses of explainability at the BBC, highlighting tensions between editorial goals and technical constraints. More broadly, there are further motivations for interpretability, including model debugging, bias detection, and trust-building \cite{brennen2020people}. Prior studies show that interpretability tools can support model debugging \cite{bhatt2020explainable}, yet even expert users such as data scientists may misunderstand or misuse explanations \cite{kaur2020interpreting}. These findings underline the importance of examining how explainability is understood and applied by engineers, who are ultimately responsible for deploying recommender systems and explanation mechanisms in practice \cite{azzopardi2024report}. 

There is a clear need for qualitative research that captures the perspectives of engineers working with news recommender systems in real-world settings \cite{Bauer2024Where}. Interview-based studies have been conducted in adjacent domains, such as evaluation practices in news \cite{Vandenbroucke2024its} and fairness and serendipity in music recommendation \cite{dinnissen2023amplifying, Binst2025what}, but work focusing specifically on explainability in news recommender systems remains limited. Our study differs from prior qualitative work~ \citep{cools2025navigating,Mitova10122023} by centering explainability within news recommender systems and by interviewing engineers across multiple news organizations in a variety of countries. 

\OurParagraph{Research questions.}
We explore the current state of personalization and explainability practices in an internationally diverse set of news organizations, guided by the overarching research question: \emph{Can explainability methods address the challenges that engineers face when building news recommender systems?} To answer this question, we investigate the following research questions:
\begin{enumerate}
    \item[\textbf{RQ1}] To what extent is personalization adopted by engineering teams at different news organizations?

    \item[\textbf{RQ2}] What are the challenges that these engineering teams face?

    \item[\textbf{RQ3}] To what extent have engineers adopted explainability for recommendations?
\end{enumerate}

\noindent\textbf{Contributions.}
First, we provide insights into the adoption of personalization practices across nine news organizations and identify key challenges faced by engineering teams. Second, we examine how explainability methods are currently used, if at all, within these organizations, and outline the reasons for their adoption or absence. Third, we offer practical guidelines for researchers and practitioners on how explainability methods can be more effectively integrated into news recommender development and deployment. 




\vspace{-1mm}
\section{Related Work}
\label{RelatedWork}
We outline prior work on journalistic values in recommendation pipelines, explainability, and studies that investigate user needs of engineers for explainability.

\subsection{Journalistic values in NRSs}
News connects people to information on current events and directly affects the functioning of democratic societies \cite{helberger2021democratic}. As online news production accelerates, news recommender systems help filter information to the right person at the right time, using a variety of approaches explored over time \cite{Wu2023Personalized}.
NRS research has developed rapidly, yet the impact on improving experiences of recommendations for a user's every day life is not so clear \cite{EkstrandMichaelD2025WNRR}, leading to disparities between what is researched and what is practised. Recent work shares lessons learned from \textit{Der Spiegel} to inform the development of more responsible NRSs through collaborations between industry and academia~\citep{bridging2026egbert}. \citet{alaqabawy2023s} argue, based on interviews with journalists, that a multi-stakeholder approach to news recommender systems is needed, with journalists as first-class users. \citet{smets2022we} conduct 11 interviews with professionals from media organizations, also finding that news recommender systems are formed through multi-stakeholder communication. 
A related study examines organizational dynamics in adopting news recommender systems across 10 media organizations, highlighting tensions between journalistic, market, and technical stakeholders~\citep{Mitova10122023}. Other work shows how journalistic values are embedded in NRSs through interviews with technical and non-technical stakeholders at two newspapers~\citep{moller2024designing}.
Diversity-aware approaches are another way of ensuring a large variety of opinions are shown to readers \cite{heitz2022benefits}. \citet{michiels2022filter} re-define filter bubbles as a decrease in any dimension of diversity. These diverse aspects can also be reflected in beyond accuracy metrics, with different metrics being relevant for different stakeholders \cite{Vandenbroucke2024its} and transparency as an important driver~\citep{grun2023transparently}. 
In this work, we provide novel insights into how news organizations embed journalistic values in their recommendation pipelines.
\vspace{-2mm}
\subsection{Explainability in recommender systems}
Explainability and transparency in recommender systems have a long history ~\cite{Herlocker2000, tintarev2015explaining}, but initially focused on explanations for users (external explainability). More algorithmic approaches to explaining recommender system decisions have recently gained popularity~\cite{Zhang2020Explainable}, and thus the evaluation of explanation quality gained popularity too \cite{piscopo2023report, Inel2023QUARE}.
Early methods in explainable AI, e.g., the post-hoc, local explanations LIME \cite{ribeiro2016should} and SHAP \cite{lundberg2017unified}, have been adapted for recommender systems in a collaborative filtering setting \cite{nobrega2019towards,zhong2022shap}. Counterfactual explanations are another post-hoc approach that focuses on explaining collaborative filtering-based recommendations  \cite{tran2021counterfactual,barkan2024counterfactual}. In contrast, \citet{ariza2024comparative} focus on generating textual explanations with LLMs.

Inherently explainable recommendation methods often leverage knowledge graphs to provide transparency. Key approaches include using reinforcement learning for path reasoning \cite{Tai2021UserCentric}, hierarchical feature learning for interpretable modeling \cite{gao2019explainable}, and specialized knowledge reasoning for news \cite{Jiang2023RCENR}.
In the content-based NRS domain, \citet{Liu2024Topic} use topic modeling to generate explanations and boost performance, while \citet{explainingMoller2025} apply integrated gradients to attribute click behavior to a specific reading history. Additionally, \citet{spisak2025segment} employ sparse autoencoders to interpret article embeddings, facilitating editorial decision-making at \textit{The Telegraph}.

Recent algorithmic advances \cite{Wardatzky2025whom} have spurred growth in user-centered explainability. \citet{szymanski2025disentangling} map explainability needs to stakeholder expertise, while \citet{integrity} demonstrate that integrity-based explanations foster appropriate trust. \citet{tran2023user} conducted a large-scale study on user needs across tourism and hospitality datasets, finding that in low-stakes scenarios, users prefer simple, positive explanations \citep{ahmad2025tell}. Conversely, in job recommendation, requirements vary significantly by stakeholder \citep{schellingerhout2023co}. Finally, \citet{waterschoot2025friends} suggest that detailed explanations in group recommenders add little value if the system is already transparent. Our work extends this earlier work to the NRS domain and investigates if news organizations have considered or are using explanations, both internally (for engineers) and externally (for their users). 

\vspace{-1mm}
\subsection{Explainability for engineers} 
\citet{bhatt2020explainable} find that engineers primarily use explainability tools for debugging and model development across domains. However, \citet{kaur2020interpreting} note that data scientists often over-trust and misuse these tools, complicating their deployment. Addressing the development process, \citet{sporsem} explores how requirements engineering strategies can elicit explainability needs to improve transparency. Building on this, \citet{habiba2025ml} interview 14 professionals to define explainability in AI-based systems and identify the practical hurdles practitioners face.

Expanding the scope to include both practitioners and end users, \citet{Liao2020questioning} investigate how different stakeholders engage with explainability. The appropriateness of explanations is highly dependent on the underlying questions being asked, and that practitioners often struggle to bridge the gap between algorithmic outputs and explanations that are meaningful and interpretable to humans. \citet{Balayn2025unpacking} examine how multidisciplinary teams, including engineers, develop trust in their organization’s LLM supply and \citet{mehrotra2025even} examine the effect of different explanation types (text, visual, and hybrid) on building appropriate trust with police officers who develop a predictive policing system. From a user interface design perspective, \citet{brachman2025towards} explore a personalized, generative-model-based user interface for model debugging and find that even within a single user group of engineers, factors such as background and task goals influence the types of explanations required.

Within the domain of news recommender systems, research focused on analyzing the explainability needs of engineers is sparse. \citet{cools2025navigating} conducts 22 interviews with mostly technical stakeholders in a single organization (BBC) and finds that although transparency and explainability are formally articulated in internal policies and AI principles, their interpretation and operationalization varies across teams. Our study extends this line of inquiry by adopting a cross-organizational perspective among different international media organizations and for different internal roles (engineers and product owners). In doing so, we present what we view as both current gaps and best practices for explanations in the field.



\section{Methods}
\label{Methods}
We conducted 15 semi-structured interviews with engineers and product owners at varying levels of seniority across nine news organizations; see Table~\ref{tab:orgs}.
Organizations were selected to have a broad international sample of institutions, i.e., private/public and news organization/aggregator. A thematic analysis was conducted on the transcribed interviews following guidelines outlined in \citep{Braun01012006}. The study was reviewed and approved by the Ethics Review Board of Eindhoven University of Technology. All participants received an informed consent form and provided consent prior to participation.

\begin{table}[t]
\centering
\caption{Organizations grouped by media type with team size (S/M or L). Team size is categorized as small-to-medium (S/M, $\leq 10$) or large (L, $> 10$). Team size is approximate, as team definitions and responsibilities vary across organizations.}

\small
\label{tab:orgs}
\begin{tabularx}{\columnwidth}{l X}
\toprule
\textbf{Media organization type} & \textbf{Organization (Team size)} \\
\midrule
Public broadcaster & NOS\footnotemark[1] (S/M), BBC\footnotemark[2] (L) \\
Newspaper/Magazine & LA Times\footnotemark[3] (S/M), DER SPIEGEL\footnotemark[4] (L) \\
Media conglomerate & JP Politiken\footnotemark[5] (S/M), DPG Media\footnotemark[6] (S/M), Schibsted\footnotemark[7] (S/M) \\
News aggregator & Ground News\footnotemark[8] (S/M), Cafeyn\footnotemark[9] (S/M) \\
\bottomrule
\end{tabularx}
\end{table}

\addtocounter{footnote}{9}
\footnotetext[1]{\url{https://nos.nl/}}
\footnotetext[2]{\url{https://www.bbc.com/}}
\footnotetext[3]{\url{https://www.latimes.com/}}
\footnotetext[4]{\url{https://www.spiegel.de/}}
\footnotetext[5]{\url{https://jppol.dk/en/}}
\footnotetext[6]{\url{https://www.dpgmediagroup.com/}}
\footnotetext[7]{\url{https://schibsted.com/}}
\footnotetext[8]{\url{https://ground.news/}}
\footnotetext[9]{\url{https://www.cafeyn.co/}}

\subsection{Interview setup}
All interviews were conducted in English and semi-structured. Interview questions were initially based on the interviews conducted by \cite{bhatt2020explainable,lucic2021multistakeholder,lucic2022towards}; we followed the guidelines in \citep{Liao2020questioning} on designing interview-based studies. All interviews were conducted by the same person in an online meeting format. Recordings were made using either Zoom or Google Meet. The duration of the interviews were planned for 30 minutes and varied between 22 and 56 minutes; see Table~\ref{tab:participants}. The questions were structured into four phases. We started with warm-up questions. These included questions such as:

\begin{bluebox}
\begin{enumerate}
\small
    \item[(1)] Can you describe your role, how long you've been with the company, and what you're working on?
    \item[(2)] What does your team do?
\end{enumerate}
\end{bluebox}

\noindent%
Next, we moved on to questions relating to the state of \textbf{personalization} at these organizations and \textbf{communicating with internal stakeholders}. These include:

\begin{bluebox}
\begin{enumerate}
\small
    \item[(3)] Do you incorporate personalized recommender systems? Why or Why not?
    \item[(4)] If so, what does that look like?
    \item[(5)] What is your recommender system development workflow?
    \item[(6)] Can you explain what type of model it is? What data is it trained on or what data do you use?
    \item[(7)] Which stakeholders do you need to communicate with within your organization?
\end{enumerate}
\end{bluebox}

\noindent%
The following stage included questions about the \textbf{challenges} that the individual and their team faces, including:


\begin{bluebox}
\begin{enumerate}
\small
    \item[(8)] What are your pain points in deploying recommender systems?
    \item[(9)] Does your model ever fail or produce strange results? How do you identify these types of results?
    \item[(10)] Think of an instance when you had to debug a model recently. What were some challenges you faced in doing this?
    \item[(11)] How did you overcome these challenges?
\end{enumerate}
\end{bluebox}

\noindent%
Lastly, we asked questions on whether \textbf{explainability}, for internal or external use, could address these challenges, following the definition of internal and external explainability in \citep{cools2025navigating}. These include:


\begin{bluebox}
\begin{enumerate}
\small
    \item[(12)] How do you try to understand your model? What do you use to try to understand your model?
    \item[(13)] If so, were these methods useful to you? Why or why not?
    \item[(14)] How familiar are you with Explainable AI research and tools?
    \item[(15)] If so, have you considered using these tools? Why/Why not?
\end{enumerate}
\end{bluebox}

\noindent%
Our goal was to obtain a broad overview of the challenges faced by news engineers. Accordingly, questions were adapted to organizational contexts, including current uses of LLMs in news and recommendation systems, as well as related applications such as article generation.

\vspace*{-1mm}
\subsection{Participants}
We recruited interviewees through our existing research network via email, LinkedIn and used snowball sampling to identify further contacts. We conducted interviews until thematic saturation was reached \cite{guest2006many}, ensuring coverage of diverse organizations, including public and private news institutions and news aggregators. Table~\ref{tab:participants} presents anonymized participant identifiers and team specifications. Team categories are derived from self-reported job titles (e.g., ``Data Scientist'' $\to$ ``Data Science''), and  abstracted team information is shown to protect participant anonymity.

\vspace*{-1mm}
\subsection{Thematic analysis}
We conducted a thematic analysis in MAXQDA 2024 \cite{maxqda24}. One researcher first cleaned and coded all transcripts, after which a second researcher independently recoded them using the same code system. The initial agreement was 84.38\% (Cohen’s Kappa $\kappa = 0.80$) \citep{landis1977measurement}, prompting a discussion that resolved disagreements in 81 coded segments and increased agreement to 96.2\%. We then used axial coding to develop themes~\citep{strauss1998basics}, iteratively grouping codes into themes and categories guided by our research questions.

\begin{table}[t]
\centering
\caption{Participants grouped and labelled by team with interview durations (in min) in parentheses.}
\label{tab:participants}
\begin{tabularx}{\columnwidth}{l X}
\toprule
\textbf{Team} & \textbf{Identifier (Duration in mins)} \\
\midrule
Data Science & D1 (30), D5 (38), D7 (56), D11 (38) \\
Product      & P2 (28), P3 (28), P10 (22), P14 (30), P15 (27) \\
Engineering  & E4 (42), E6 (31), E8 (27), E9 (27), E12 (30), E13 (33) \\
\bottomrule
\end{tabularx}
\end{table}



\vspace{-1mm}
\section{Results}
\label{Results}
In this section, we report the results to our stated research questions. Participants are referred to using the identifier codes in Table \ref{tab:participants}. 

\begin{figure*}
    \centering
    \includegraphics[width=\linewidth]{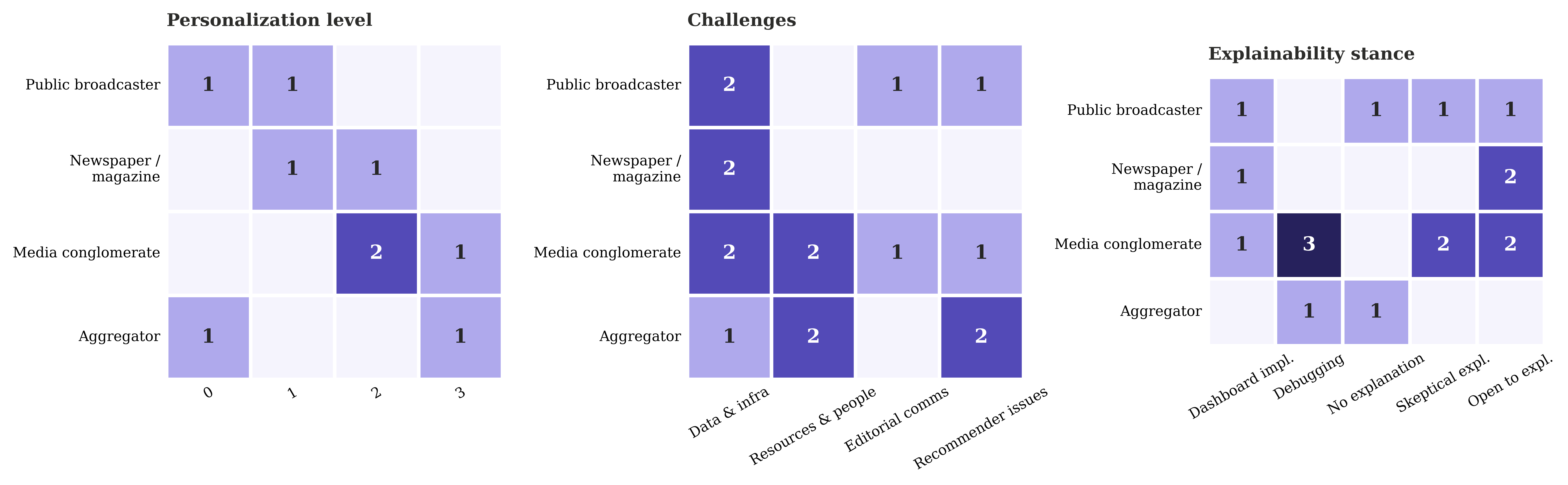}
    \caption{Cross-organizational overview of key findings across RQ1–3. Each cell reports the number of organizations of a given type exhibiting that characteristic. (Left) Personalization levels range from 0 (no personalization) to 3 (pro-personalization). (Middle) Challenges reflect infrastructural and organizational barriers to recommender system deployment. (Right) Explainability stances capture how organizations approach or resist algorithmic transparency.}
    \label{fig:cross_org_overview}
    \Description[Three heatmaps show personalization level, deployment challenges, and explainability stance by organization type]{Three side-by-side heatmaps compare four organization types, public broadcaster, newspaper or magazine, media conglomerate, and aggregator, across personalization level, challenges, and explainability stance. In the personalization level heatmap, scored 0 to 3, public broadcasters cluster at the low end with one organization each at 0 and 1. Newspapers and magazines are at 1 and 2, one organization each. Media conglomerates are highest, with two organizations at 2 and one at 3. Aggregators are split, with one organization at 0 and one at 3. In the challenges heatmap, data and infrastructure issues are the most common challenge overall, reported by two public broadcasters, two newspapers or magazines, and two media conglomerates. Media conglomerates also report two organizations facing resource and people challenges, plus one each facing editorial communication and recommender system issues. Public broadcasters report one organization each facing editorial communication and recommender system issues. Among aggregators, one organization reports facing data and infrastructure challenges, two facing resource and people challenges, and two facing recommender system issues; none report editorial communication challenges. In the explainability stance heatmap, media conglomerates show the widest spread, with three organizations focused on debugging, two skeptical of explanation, two open to explanation, and one each using dashboard implementations. Public broadcasters have one organization each in dashboard implementation, no explanation, skeptical of explanation, and open to explanation. Newspapers and magazines have one organization using dashboard implementation and two open to explanation. Aggregators have one organization focused on debugging and one reporting no explanation, with no organizations open to explanation.}
\end{figure*}

\vspace*{-1mm}
\subsection{To what extent is personalization adopted?}
\label{Results:personalization}
The first part of the interviews pertained to understanding if and how personalization exists in NRSs. We group organizations into those that use personalization in news recommendation and those that do not. We then discuss reasons why organizations are adopting or avoiding personalization. Lastly, we analyze our results from a cross-organizational perspective, using the aforementioned groupings to define personalization levels in Figure~\ref{fig:cross_org_overview} (left).

\OurParagraph{Organizations using personalization.}
Most organizations, seven out of the nine interviewed, stated that personalization is a part of the wider user experience, on the website, in an app, or in a personalized newsletter.  Implementations of personalization pipelines differ. E.g., P3 indicated their organization has a unified approach: ``\textit{we can use the same recommendation engine on the front page as we can in a box underneath an article in a widget}.'' Other organizations have a segmented approach ``\textit{For every purpose there is a different algorithm}'' (D7). In between these two perspectives, three participants mentioned that their personalized news pipelines mainly contain two or three types of approaches: content-based news recommendation, collaborative filtering, and/or content-to-content based recommendation which is non-personalized. E.g., D1, P2 and E8 mentioned a personalized approach and a non-personalized approach, with E8 stating ``\textit{We currently have two flavors of personalization. We have the content-based personalization [\ldots] And then we have a collaborative filtering flavour}.'' Similarly, D1 noted that ``\textit{one is personalized and one is not personalized, but it's based on what is performing well, and showing that content to more users. We've had really good success with that latter bucket.}'' Interestingly, P2 also mentioned non-personalized approaches already having a positive impact for the organization: ``\textit{And then a lot of `read more' at the bottom of an article [\ldots] But even that was a big win in automating that for the news brands}.'' In the case of D11, the recommender system is outsourced, using ``\textit{out of the box solution}'' that is``\textit{very much content-based with some business rules for re-ranking and filtering.}''

\OurParagraph{Organizations not using personalization.}
A smaller group of organizations indicated, at the time of the interview, not to incorporate any personalization in their pipeline. E.g., similarly to the content-to-content method mentioned by D1 and P2, P10 stated that ``\textit{when a user opens a certain article, we're trying to find articles that are similar to that article that they're reading right now [\ldots] So, recommendations are also the same for everyone who opens that article. It's not personalized}.'' Another organization recently moved away from personalization, with E9 stating ``\textit{We did that for six months. We actually recently just removed it [\ldots] That is why news recommendation is a very interesting challenge, or a product for news because it goes against the ethos of our organization in a way}.'' As an alternative, E9 mentioned a larger focus on categorizing articles and allowing users to filter based on their interests and that ``\textit{everybody sees it as a newspaper. Everybody gets the same treatment}.''

\OurParagraph{Reasons for moving towards personalization.}
Participants’ responses frequently prompted follow-up questions about the rationale behind organizational decisions and whether investments in personalization would continue. E6 and D7 noted that their organizations are moving forward with automating the front page. D7 described how a previously manual newsletter, where ``\textit{everyone got the same}'' became partially automated after experimenting with personalization. This was particularly effective for topics the editorial team ``\textit{simply stayed out}'' of, such as cryptocurrency, which was not considered relevant for most users but ``\textit{nice to get}'' for a minority, marking ``\textit{the first step of replacing humans}'' (D7). Other organizations fall somewhere in between automation and editorial control (P2, P3, D11). P3 eluded to the balance between editorial and personalized news with ``\textit{we're trying to check off sort of all the possible contexts while still safeguarding the journalistic mission because we aren't YouTube, we aren't Instagram, we aren't TikTok. We are serving news to the public and still have editors who at any time decide what's most important that the public should know now}.''

\OurParagraph{Reasons for moving away from personalization.}
Whilst most organizations have personalized approaches incorporated in their products, some deliberately choose to avoid personalization. P10 stated ``\textit{one of the reasons is we're just not a tech company}'', and continued with ``\textit{The other reason why we're keeping it simple is because we don't want to gather a lot of personalized data}.'' E9 cited the issue of echo chambers as a reason not to personalize, however, mentioning that they may bring it back ``\textit{Maybe in the future recommendations will come back [\ldots] For news, it's kind of a slippery slope}.'' Even for organizations that fall somewhere in between this spectrum, there are questions on how much to invest in personalized experiences. D1 mentioned ``\textit{We're also at a bit of a fork in the road of whether we're going to stick with that [\ldots] do we want to really invest a lot more in this system and make those kinds of improvements?}'' Another organization reported prioritizing personalization for its video platform, with news taking a back seat. D5 explained that users may view the platform similarly to a Netflix subscription, where payment implies personalized value, making personalization ``\textit{one of the top priorities}'' for their video platform.

Zooming out with Figure~\ref{fig:cross_org_overview} (left), we find that personalization in news is less widely adopted by public broadcasters than by other types of media organizations. This is unsurprising, as public broadcasters typically lack access to user data comparable to that of privately funded organizations, where logins are often tied to subscription-based funding models. More surprising, however, is the variation within privately funded organizations: one aggregator and one newspaper/magazine fall at the lower end of personalization, while the others exhibit higher levels of personalization.
\vspace*{-1.5mm}
\subsection{What are the challenges?}
\label{Results:challenges}
Drawing on questions 8–11 from Section~\ref{Methods}, we identify four categories of challenges related to personalization and discuss their impact on different media types.

\OurParagraph{Data, backend and infrastructure.}
When asking participants about their challenges, the overwhelming majority pointed to data, backend or infrastructure as a key concern (P3, E4, D5, E6, D7, P10, D11, P15). E.g.,  D7 said that ``\textit{most of the time your backend land is pretty disjoint from your data land}'' and D11 stated that these challenges made it ``\textit{a little bit hard to debug the system}.'' E4 indicated that transitioning from one infrastructure to another causes issues, with their recsys team becoming ``\textit{a centralised unit}'' while building ``\textit{a huge new data structure}.'' P10 mentioned being bottlenecked by old infrastructure. P3 emphasized access to data and subscriber retention, noting that ``\textit{the GDPR makes it really hard to get good data that would benefit society. We believe that it would benefit society if more people would have a newspaper subscription}.'' Improving retention is also a concern for P14: ``\textit{our biggest pain point is that we want to use recommendations to keep people on the platform, and we can optimize and measure relatively short term results like clicks and plays}.'' 

\OurParagraph{Resources and people.}
Another stated challenge by participants (P3, E4, D7, E9) was access to resources and people. E9 mentions the ``\textit{big computational challenge}'' of running recommender systems and D7 stated that their``\textit{current data science team, it’s four people. We don’t have this in-depth knowledge. We don’t have a specialist. Because at this point, we need generalists\footnote{Hiring constraints force a focus on generalists, even though new technology requires specialists.} who can make changes on the product and not only focus on optimising specific parts}.'' 

\OurParagraph{Communicating with editorial teams.}
Communication and talking to the newsroom often came up as a challenge (P2, D5, P10, P14). E.g., P2 referred to informing newsrooms about articles that were generated and how that is difficult because they are ``\textit{quite a big company with also a lot of silo work within those news brands, not everyone is always up to date}.'' P14 mentioned editorial reviewing changes, which can also slow things down, and P10 mentioned they ``\textit{always want to make sure that people like our colleagues who are editors know that we are not planning to replace their work with AI}.''

\OurParagraph{Issues with the recommender system.}
Lastly, several participants reported pain points with how the recommender system actually worked (D5, E9, E12, P14). Some argued that different products require different recommender systems (D5) and indicated that they are currently not using ``\textit{cross-product interactions}.'' In contrast, E9 mentioned diversifying recommendations and how this is particularly challenging with news where recommendations should not only focus on what is trending. E12 mentioned issues with news categories like sports where there are either ``\textit{too many}'' or  ``\textit{too few}'' articles for different types of user. P14 found complex business rules challenging, stating that ``\textit{we’ve got lots of business rules to constrain our recsys outputs. So, for instance, if you liked a politics show isn’t a thing we’d ever show because if you like, a certain political persuasion isn’t what we want to promote. So we’d have a business rule. And if you liked, well, that would say,
you know, here are the recommendations. Now strip out any political titles, but then a different recommender might have a very similar rule, but it will be written in a different way. Not one is right or wrong, but just different teams at different points}.'' The cold start problem in news came up with some participants as a key challenge (D7, E8, P14). Other participants mentioned solutions such as LLMs (P3) and using non-personalized bandit algorithms for exploratory recommendations (P15). 

From a cross-organizational perspective (Figure \ref{fig:cross_org_overview}, middle), data, backend, and infrastructure emerge as universal challenges for engineers. Resource and staffing constraints might be expected to be widespread, but they are reported only by aggregators and media conglomerates. Challenges related to recommender systems are most prevalent among organizations with higher levels of personalization.

\subsection{To what extent is explainability adopted?}
\label{Results:explainability}
This study assessed how engineers adopt explanations, allowing participants to define the concept in their own terms. No single organization explicitly stated to use explainability or interpretability algorithms for any purpose. This does not mean that organizations do not have a means to debug models, to communicate with the newsroom or to be transparent on their websites or apps. We identified five categories of explanation (related) practices and vision (see Figure~\ref{fig:cross_org_overview} (right)): two focus on internal explainability (dashboards and debugging), the other three about external explanability (no explainability, skeptical about explainability, and open to explainability in the future).

\OurParagraph{Dashboards for the newsroom.}
No organization explicitly uses explainability tools but three mentioned having a dashboard for internal use, specifically to communicate and monitor changes to the recommender system. P3 mentioned ``\textit{the editors that we bring in the projects, they get to see everything. We try to explain how the model works, the feature significance. We have different tools where we can look up articles and say alright I understand that you would think that this article would perform much better but when we look at the data and the features on this we see that it didn't really attract the young users as you had expected.}'' P15 and D5 interact with editorial teams through metrics and dashboards: ``\textit{We use a combination of metrics [\ldots] including diversity metrics, coverage metric [\ldots] but very often we have an internal review with bespoke visualisation tools, which are used to show recommendations to editorial. And they're very helpful to spot inappropriate pairings}'' (D5). P15 noted ``\textit{now I can run a dashboard and a report, and I can show you that, in fact, we are not doing what you asked and that we are representing our values inside of the system with real data, with real dashboards. This was very compelling}.'' 

\OurParagraph{Debugging approaches.}
When asked questions 12--16, some participants mentioned debugging approaches and how they understand their model. E6 mentioned explainability for ``\textit{Just for the other IT people}'' and that ``\textit{we make an effort like in the end we want to transform all the scores so they’re between 0 and 1.\footnote{Here, the 0 and 1 refer to the normalized bounds of a model's output score.} Because then when you see the raw data, you can tell what a high score \& a low score is. So that’s one type of explainability. It’s just useful for the other IT people.}'' E8 and E9 do manual checks to understand where things went wrong, such as ``\textit{we just select a few users and we do some checks with the models}'' (E8) and ``\textit{one of the ways we debug is to look at what was the previous user's behavior and what other users who are closest to this user, on what their behaviors are}.'' When referring to engineers using feature attribution for debugging in their team, P3 said ``\textit{they probably use everything};'' usage is currently unknown, but engineers are free to use the tools as needed. 

\OurParagraph{No external explainability.}
Two organizations stated that they would not use explanations externally, although for very different reasons. D7 stated ``\textit{I think we discuss it twice a year. I'm not fully convinced this is the thing that our users want. [\ldots] This is shared by the product team. So we do discuss and it does come up, I know the techniques are there. We could do something about it. But it's not a way that we can differentiate [\ldots] from a business perspective. But I also think that there's only a small group of users who actually would get the benefit from it.''} The motivation for not incorporating explanations is linked to the history in recommender systems with user controllability and customizability, further mentioning ``\textit{Every news site tried it out. Every news site disbanded it, that users get this full control. Because the moment that you give them control, it means that it makes it much harder for you. If you want to take control, to push something that you think is important. Either from a journalistic perspective or from something that needs to be pushed. It's the same here}.'' In contrast, E13's reason for not including explanations is that if you avoid personalization, you also avoid black box scenarios where explanations are necessary ``\textit{I think by designing it the way we did now, explainability is not like a hard requirement.}'' 

\OurParagraph{Skeptical about external explainability.} Other engineers acknowledged that explanations have some concerns (P2, E4, E6, P10, E12). P2 mentioned that there may be fear from newsrooms with complaints from readers, stating ``\textit{So it’s like don’t poke the bear if there’s no reason to consider people being unhappy with the lists}'' where lists are the algorithmic outputs such as ranked news feeds or content suggestions. E4 and P10 questioned whether this is what users even want:  ``\textit{Within the first year [of deploying the recommender system] we had one person contact us. And that was some student that was interested in it. So, I’m not sure how many people actually care about it}'' (E4) and ``\textit{information overload if you place an explanation with every recommendation}'' (P10). E6 and E12 mentioned risks with explanations stating ``\textit{It can always be done badly}'' (E6) and ``\textit{it can be manipulated a lot}'' (E12). 

\OurParagraph{Open to external explainability in the future.}
The last group of participants seemed open to the idea of incorporating explanations to readers in the future (D1, E6, D11, P14). D1 mentioned a potential ``\textit{next project is like a specific module that’s like `because
you read'}.'' Again, we see a connection with personalization and a growing need for explanations, with E6 stating ``\textit{as you take more and more [of the front page], this explainability is in focus, both for our own sakes so we can know what’s going on, but also for the end user}.'' For P14, generic explanations have already been implemented, such as ``\textit{based in the area},'' but more personalized explanations have not been tested. For D11 there is a lot of interest within their team, mentioning a researcher that ``\textit{is very interested in transparency. So communicating the fact that it is personalised to the users. So we will do some experiments with this on the website this year. And we are also interested in a controllability. And that is a little bit related to explainability. It depends a little bit on your definition of explainability}.'' For most participants, a common reason why explainability has not yet been incorporated, even if they are positive about it, is a prioritization for current projects, e.g., optimizing current recommender systems (D1, E6, D11, P14), and internal challenges, e.g., data infrastructure (D11, P14) and lack of time (E6). 

A cross-organizational view of explainability adoption (Figure~\ref{fig:cross_org_overview}, right) reveals broad receptiveness, with only a subset of organizations skeptical of its utility. A clear pattern emerges among media conglomerates: all cite debugging as a key lens for interpreting recommendation results, reinforcing the earlier observation that personalized recommendation pipelines are prevalent in this sector.



\vspace*{-2mm}
\section{Overarching Themes}
\label{Themes}
Three themes emerged: newsroom influence on personalization, a lack of stakeholder explanations, and unexpected findings outside of the original scope.

\vspace*{-1mm}
\subsection{Editorial control}
In all organizations except news aggregators, editors play a central role in defining brand values, shaping NRS design, and influencing system transparency. Even the two aggregators we interviewed apply business rules grounded in their organizational values. Communicating these editorial values to the newsroom can be challenging for engineers (see Section~\ref{Results:challenges}).
Despite this, news engineers realize that personalization in news has valid constraints due to the journalistic values (P3: ``\textit{Editors decide what the public should know}'') and privacy concerns of the organizations (P10: ``\textit{don't want to gather personal data}''), which are expressed by editorial teams. Because of these constraints, personalization may not need to be as developed as in other domains such as e-commerce and movie recommendation (P14: ``\textit{we rely on editorial stakeholders for a lot}'').

Values that are embedded into the fabric of the organization are reflected through the business rules (D11: ``\textit{we have developed them together with the editors}''). Almost all participants mentioned them as a crucial part of developing their recommender systems (D1, P2, P3, D5, E6, D7, E9, D11, E12, E13, P14, P15). An example of editorial control through business rules is a temporal cutoff (D1: ``\textit{So weather stories [\ldots] they have a very clear expiration. And we do have mechanisms for kind of editorial control over that}''). Another example of business rules in recommendations are sensitive news topics. D5 mentioned an anecdote on issues with the editorial and a content-based news recommender system: ``\textit{[there were] arson attacks at a synagogue in New York [\ldots] the recommender linked that article immediately with something about Holocaust Memorial Day. But I don't want to link everything that's related to Jewish people to the Holocaust. And so editorial told us, well, this might be inappropriate}.'' 

Business rules are not just a part of the recommender design, but they clearly tie in with the results on explanations in Section~\ref{Results:explainability}, where dashboards in the newsroom can be used to monitor and possibly develop business rules (P3, D5, P15). Identifying problems or bugs in recommendations can also motivate and inform new business rules. For example, E12 mentions where they are ``\textit{playing around with the `most read' list because there is actually a lot of things that you can optimize on [\ldots] usually they just say you should count [the last] 24 hours, but if you take at 4 [pm] today, then the most popular article is the one that was popular yesterday}.''

Another sub-theme across participants was the importance of diversity in recommendation (D1, E4, D7, E9, D11, E12, E13, P14, P15). Diversity emerges as a challenge, a business rule, an evaluation criterion, and a core objective of the recommender system, with multiple dimensions shaping what counts as a diverse recommendation. E.g., E13, who does not personalize their NRS, said that ``\textit{even if it's not the perfect recommendation, it's still kind of fine because we generally also just want to present the user with more content that they might be interested in}''. P15 related a lack of diversity to news fatigue, i.e., too many articles about wars may lead to ``\textit{a decline in the visits per user}.'' There could also be more diversity in article stances on the same topic or from different sources (D7, E9). 

\vspace*{-1mm}
\subsection{Explainability vs.\ Understanding}
Even without formal explainability tools, engineers still seek to understand and communicate how their recommender works. As Section \ref{Results:explainability} shows, most debugging is manual and unsystematic. More analytic monitoring could involve feature attribution, counterfactual explanations, or examining article embedding spaces, an approach some engineers explicitly described when trying to understand the high-dimensional space of article text (D1, P2, E9, P10, D11, E12). E12 said ``\textit{sport articles are very easy if you take the embeddings and then you visualize it, then sports is very easy to classify but everything else is kind of all over the place}.'' 
Engineers do not always find manual checking problematic, but some are more eager to improve: ``\textit{I don't think we have a method to entirely capture all things that could go wrong. And I'm not sure if it even exists. Because if it does, it would be great to know about it}'' (P10). 

Interestingly, when talking about explainable AI, many engineers pivoted to the `old school' recommender systems explanations for end-users that could be found on Netflix (E6: ``\textit{Sort of like Netflix [...] we have talked about that}'') or Facebook (E12: ``\textit{explain it by `your friends are also looking at'}'') \citep{tintarev2015explaining}. Surprisingly, algorithmic approaches were not favored, despite aligning better with technical stakeholders.  We see from discussions on bugs in recommendations (Section~\ref{Results:challenges}) and investigations into the article embedding space that there is a desire for tools that can better equip them to understand their models better and not necessarily to explain to users.

\vspace*{-1mm}
\subsection{LLM Adoption}
While outside our original research questions on personalization and explainability, LLM adoption emerged consistently enough across interviews to warrant separate note: many organizations are adopting LLMs, for article generation or editing (P2, E4, E6, D7, E13), generating meta-data (D1, D7) or as an internal tool (P2, E4, E6, E12). Regarding LLMs as an internal tool, E4 said for now ``\textit{the target customers are the journalists and the editors}.'' E12 was skeptical of the tool for model debugging: ``\textit{Everybody says that we should use the chatbot and I really tried and it took me on so many goose chases [...] it made mistakes that took me longer time to find than it probably would have done if I hadn't tried to get help in the first place}.''

LLMs are used to generate articles, reports, and weather summaries (P2, E4). While boosting efficiency, e.g., in automated newsletters (D7), this has also led to job displacement. E6 mentioned that ``\textit{we have made our own sort of ChatGPT which has resulted in some proofreaders which have been fired. This was not very popular}.'' With article text generation, there are concerns about their alignment with editorial values. E.g., E13 stated that their organization was ``\textit{very hesitant with using the word genocide, regarding what's going on in Israel and Gaza and there's lots of critique. So sometimes, if you ask them [LLMs] to give a certain summary and it uses a word that's not the word that you would want to be responsible for. [...] we as an organisation are responsible for what the LLM writes at the end of the day and it's easier to hold people accountable than it is with a chatbot}.'' E13 mentioned using Model Context Protocol (MCP) servers to improve the article position on services such as ChatGPT: ``\textit{so besides search engine optimization, we also have some sort of chatbot search optimization}.''

\vspace*{-2mm}

\section{Discussion} 
\label{Discussion}


\OurParagraph{Key insights \& implications.}
Our interviews reveal a complex landscape in which personalization and explainability are unevenly adopted and difficult to implement. Three key insights stand out: (i)~Contrary to much algorithmic research, personalization is neither universally desirable nor always feasible; it clashes with editorial control, user privacy, and journalistic missions. (ii)~Interviews exposed substantial operational and infrastructure challenges, including limited team sizes (D7: ``\textit{four people \ldots\ we don't have a specialist}''), and competing priorities that constrain what organizations can implement. This creates a significant gap between academic methods and what news organizations can practically deploy. (iii)~When organizations invest in transparency, they mostly build internal dashboards rather than reader-facing explanations. Explainability tools thus mainly support internal debugging, echoing~\citep{bhatt2020explainable} and extending it by suggesting that explainability research should target multi-stakeholder communication, not only end-user transparency.

\OurParagraph{Relation to prior research.}
Our insights align with \citet{cools2025navigating}'s finding at the BBC that explainability remains aspirational rather than operational. Our cross-organizational perspective reveals that these challenges are \emph{systemic}, not organization-specific.
Further examining this from a cross-organizational perspective, we find that, overall, the differences between public and private organizations are not that large. There is a spectrum, with public organizations valuing data privacy more and some private organizations having more financial incentives for automating jobs. But both types of organization embed their values in similar ways and have similar views on explainability.
Consistent with \citep{Mitova10122023}, we observe tensions among journalistic, market, and technical stakeholders, and extend it by showing how these tensions manifest in explainability decisions. Our findings also extend \citep{bhatt2020explainable,habiba2025ml}, revealing an additional critical function in news: bridging engineering and editorial teams.

\OurParagraph{Guidelines.}
Returning to the wider question of \textbf{whether explainability methods could address the challenges that engineers face when building news recommender systems}, we believe they could and should play a role in communication with the newsroom and with monitoring bugs in a system (i.e., explainability-for-maintainers rather than explainability-for-end-users, per the scrutibility definition above). But there are organizational challenges that cannot be solved in this way, like access to better data, moving to new infrastructures, and a lack of financial resources and people.  Based on our analysis, we recommend: (\textbf{i})~Start with internal explainability for editorial communication through dashboards showing recommendation behavior and diversity metrics (as demonstrated by P3, D5, P15), rather than complex end-user explanations, given their uncertain benefits (E4: ``\textit{only one user inquiry in a year}'') and potential risks (E6: ``\textit{can be done badly}"). (\textbf{ii})~Adopt more systematic debugging approaches instead of score normalization, manually testing user profiles and embedding visualizations, instead of investing in more sophisticated methods. And (\textbf{iii})~Develop explainability requirements alongside business rules, creating traceability for when rules trigger.

\OurParagraph{Potential impact.}
Our work can have significant impact across research, industry, and policy. For researchers, it bridges the gap between explainability research and industrial practice by revealing what actually works in production environments and identifying real-world barriers that can guide future tool design. For practitioners, it enables organizations to learn from peers facing similar challenges, offers actionable guidelines that can reduce implementation costs, and helps teams make informed decisions about investing in explainability methods. Beyond immediate applications, this research can inform regulatory discussions around transparency requirements by grounding them in practical realities, influence industry best practices for responsible NRSs deployment, and affect millions of news consumers by improving how personalized systems are developed and made transparent.

\OurParagraph{Limitations.}
Our sample focuses on European and North American organizations under specific regulatory frameworks; findings may not generalize globally. We primarily captured technical perspectives; future work should incorporate editorial and reader views. Cross-stakeholder research on how engineers, editors, journalists, readers, and regulators perceive and value explainability could inform more nuanced design decisions.



\vspace*{-1mm}
\section{Conclusion}
\label{Conclusion}
We have conducted 15 semi-structured interviews with news engineers and product owners to investigate the extent to which personalization and explainability are currently developed in different news organizations. Through these interviews, we have found that personalization adoption is neither universal nor straightforward. Organizations make deliberate choices to limit or avoid it based on editorial values, privacy concerns, and resource constraints. Seven out of nine organizations use some personalization, but most employ hybrid approaches balancing algorithmic recommendation with editorial control. No organization explicitly uses algorithmic explainability tools, despite interest from some teams. Where explainability exists, it focuses on internal stakeholders rather than end users, with external explanations viewed as unnecessary, risky, or low-priority. 

For explainability to realize its potential in news recommendation and personalization, the RecSys community has both an opportunity and a responsibility to design methods that recognize constraints and provide lightweight, production-ready solutions that integrate into existing workflows. This work represents a step toward that goal, but much work remains to be done.
A particularly important next step in this direction would be to expand the study to journalistic, market, editorial and end-user stakeholders.


\begin{acks}
     We thank all participants for their time and valuable insights. We are also grateful to everyone who assisted with participant recruitment, and thank Hannes Cools, Maria Heuss and the AI, Media and Democracy Lab for their feedback.
     This research was (partially) supported by the Dutch Research Council (NWO), under project numbers 024.004.022, NWA.1389.20.\-183, and KICH3.LTP.20.006, and the European Union under grant agreement No. 101201510 (UNITE).
     Views and opinions expressed are those of the author(s) only and do not necessarily reflect those of their respective employers, funders and/or granting authorities.
\end{acks}

\bibliographystyle{ACM-Reference-Format}
\balance




\appendix

\end{document}